\documentclass [12pt]{article}
\usepackage {graphicx}
\begin{document}

\begin{center}

\textbf{Scattering of electromagnetic waves by many thin cylinders: 
theory and computational modeling}
\end{center}

\bigskip

 A. G. Ramm$^{1}$ M. I. Andriychuk$^{2}$%$, A. G. Ramm$^{2}$

%$^{2}$Institute for Applied Problems in Mechanics and Mathematics, NASU,

%Naukova St., 3B, 79060, Lviv, Ukraine

%E-mail: andr@iapmm.lviv.ua

$^{1}$Mathematics Department, Kansas State University,

Manhattan, KS, 66506, USA

E-mail: ramm@math.ksu.edu

$^{2}$Institute for Applied Problems in Mechanics and Mathematics, NASU,

Naukova St., 3B, 79060, Lviv, Ukraine

E-mail: andr@iapmm.lviv.ua

\bigskip

%%%%%%%%%%%%%%%%%%%%%%%%%%%%%%%%%%%%%%

{\bf Abstract}
\bigskip

Electromagnetic (EM) wave scattering by many parallel infinite cylinders 
is 
studied asymptotically as $a \to 0$, where $a$ is the radius of the 
cylinders. 
It is assumed that the centers of the cylinders $\hat {x}_{m} $ are 
distributed 
so that $\mathcal{N}(\Delta ) = {{\rm ln}(\frac{{1}}{{a}}}){\int\limits_{\Delta} 
{N(x)dx[1 + o(1)]}} $,
 where $\mathcal{N}(\Delta )$ is the number of points $\hat {x}_{m} = 
(x_{m1} ,x_{m2} )$ in an arbitrary open subset of the plane $xOy$, the 
axes of cylinders are parallel to $z$-axis. The function $N(x) \ge 0$ is 
a given continuous function. An equation for the self-consistent 
(limiting) field is derived as $a \to 0$. The cylinders are assumed 
perfectly conducting. Formula for the effective refraction coefficient 
of the new medium, obtained by embedding many thin cylinders into a 
given region, is derived. The numerical results presented demonstrate 
the validity of the proposed approach and its efficiency  for solving 
the many-body 
scattering problems, as well as the possibility to create media 
with  negative refraction coefficients.

\bigskip

%%%%%%%%%%%%%%%%%%%%%%%%%%%%%%%%%%%%%%
Key words: EM wave scattering by many thin cylinders; asymptotic 
solution; refraction coefficient; effective medium theory; 
nanowires; computational 
modeling

\bigskip

%%%%%%%%%%%%%%%%%%%%%%%%%%%%%%%%%%%%%%%
{\bf 1. Introduction}
%%%%%%%%%%%%%%%%%%%%%%%%%%%%%%%%%%%%%%%

\bigskip

Wave scattering by many thin cylinders (nanowires) is important 
because of its many applications in chemistry [9], [22], medicine [23], 
optics [4], 
nanotechnology [5], etc. Analytical formulas for solving 
electromagnetic (EM) wave scattering problem by many thin cylinders were 
derived in [19], and the results from [19] are used in this paper. 

There is a large literature on 
EM wave scattering by  arrays of parallel cylinders (see, for 
example, [7], [8]). Our approach has the following novel features: 

- the cylinders are thin, that is,  they have small radius $a$,  
$ka \ll 1$, where $k$ 
is the wavenumber of the medium outside of the cylinders; this 
allows one to obtain a rigorous asymptotic solution of the
wave scattering problem by many thin cylinders;

- the solution to the wave scattering problem is considered 
also in the limit $a \to 0$ when 
the number $M = M(a)$ of the cylinders tends to infinity at a suitable 
rate and the distance $d$ between neighboring cylinders is much greater than 
$a$, but there can be many small cylinders on the wavelength, so that
the multiple scattering effects are essential; these effects are taken 
into account rigorously; both analytical and numerical methods for
solving wave scattering problem on these thin cylinders (nanowires)
are proposed and tested numerically;

- the theoretical results obtained is a basis for a method for creating 
a new medium with negative refraction 
coefficient $n^{2}(x)$; this new medium is obtained by embedding many 
small perfectly conducting cylinders into a given (initial) medium.

This work continues the earlier investigations in 
[10]-[21], and the numerical modeling presented in [1]-[3].

Let $D_{m} ,1 \le m \le M$, be a set of non-intersecting domains on a 
plane $P$, which is  $xOy$ plane. Let $\hat {x}_{m} \in D_{m} 
,\hat {x}_{m} = (x_{m1} ,x_{m2} )$, be a point inside $D_{m} $, $S_m$ 
be the boundary of  $D_{m} $, and 
$C_{m} $ is the cylinder with the cross-section $D_{m} $ and the axis, 
parallel to $z$-axis,
 passing through $\hat {x}_{m} $. We assume that $\hat {x}_{m} $ is 
the center of the disc $D_{m} $ if $D_{m} $ is a disc of radius $a$. 

Let us assume that the cylinders are perfect conductors and 
$a = 0.5{\rm diam} D_{m} $. Our "smallness" (thinness) assumption is 

\begin{equation} \label{eq1} ka \ll 1, \end{equation} 

\noindent where 
$k$ is the wavenumber in the region exterior to the union of 
the cylinders. 

We assume that the thin cylinders are distributed according to the 
following law:

\begin{equation}
 \label{eq2} {\mathcal N}(\Delta ) = \ln 
{\frac{{1}}{{a}}}{\int\limits_{\Delta} {N(\hat {x})d\hat {x}[1 + 
o(1)],\,\,\,\,\,\,\,a \to 0}} ,
 \end{equation}

\noindent
where ${\mathcal N}(\Delta ) = {\sum\limits_{\hat {x}_{m} \in \Delta}  {1}} $  is 
the number of the cylinders in an arbitrary open subset of the plane $P$, 
$N(\hat {x}) \ge 0$ is a continuous function, which can be chosen as we 
wish. The points $\hat {x}_{m} $ are distributed in an arbitrary large but 
fixed bounded domain in the plane $P$. We denote by $\Omega $ the union of 
domains $D_{m} $, by $\Omega '$ its complement in $P$. The complement in 
${\rm R}^{{\rm 3}}$ of the union $C$ of the cylinders $C_{m} $ we denote by 
$C'$.

The EM wave scattering problem consists of finding the solution to 
Maxwell's equations

\begin{equation}
\label{eq3}
\nabla \times E = i\omega \mu H,
\end{equation}

\begin{equation}
\label{eq4}
\nabla \times H = - i\omega \varepsilon E,
\end{equation}

\noindent
in $C'$ such that

\begin{equation}
\label{eq5}
E_{t} = 0\,\,{\rm o}{\rm n}\,\,\partial C,
\end{equation}

\noindent
where $\partial C$ is the union of the surfaces $C_{m} $, $E_{t} $ is the 
tangential component of $E$, $\mu $ and $\varepsilon $ are constants in 
$C'$, $\omega $ is the frequency, $k^{2} = \omega ^{2}\varepsilon \mu $. 
Denote by $n_{0}^{2} = \varepsilon \mu $, so $k^{2} = \omega ^{2}n_{0}^{2} $. 

Let us look for the solution to problem (3)-(5) of the form

\begin{equation} 
\label{eq6} E(x) = E_{0} (x) + v(x),\,\,x = (x_{1} 
,x_{2} ,x_{3} ) = (x,y,z) = (\hat {x},z), 
\end{equation}

\noindent
where $E_{0} (x)$ is the incident field,  $v(x)$ is scattered field 
satisfying the radiation condition

\begin{equation}
\label{eq7}
\sqrt {r} ({\frac{{\partial v}}{{\partial r}}} - ikv) = o(1),\,\,\,r = 
(x_{1}^{2} + x_{2}^{2} )^{1 / 2},
\end{equation}

\noindent
and we assume that

\begin{equation}
\label{eq8}
E_{0} (x) = e^{i\kappa y + ik_{3} z}e_{1} ,\,\,\,\kappa ^{2} + k_{3}^{2} = 
k^{2},
\end{equation}

\noindent
where $\{e_{j} \},\,j = 1,2,3,$ are the unit basis vectors along the 
Cartesian coordinate 
axes $x,y,z$. We consider EM waves with $H_{3} : = H_{z} = 0$, i.e., 
$E$-waves, or $TH$-waves,

\begin{equation}
\label{eq9}
E = {\sum\limits_{j = 1}^{3} {E_{j} e_{j}}} ,\,\,\,H = H_{1} e_{1} + H_{2} e_{2} 
= {\frac{{\nabla \times E}}{{i\omega \mu}} }.
\end{equation}

It is proved in [19] that the components of $E$ can be expressed 
by the formulas:

\begin{equation}
\label{eq10}
E_{j} = {\frac{{ik_{3}}} {{\kappa ^{2}}}}u_{x_{j}}  e^{ik_{3} z},\,\,\,j = 
1,2,\qquad E_{3} = ue^{ik_{3} z},
\end{equation}

\noindent
where $u_{x_{j}}  : = {\frac{{\partial u}}{{\partial x_{j}}} }$; the 
function $u = u(x,y)$  solves the problem

\begin{equation}
\label{eq11}
 (\Delta ^{2} + \kappa ^{2})u = 0\,\,{\rm i}{\rm n}\,\,\Omega ',\quad  \kappa 
^{2}:=k^2-k_3^2,
\end{equation}

\begin{equation}
\label{eq12}
{\left. {u} \right|}_{\partial \Omega }  = 0,
\end{equation}

\begin{equation}
\label{eq13}
u = e^{i\kappa y} + w,
\end{equation}

\noindent
and $w$ satisfies the radiation condition (\ref{eq7}). 

One can check (see [19]) that the unique 
solution to (\ref{eq11})-(\ref{eq13}) is given by the formulas:

\begin{equation}
\label{eq14}
E_{1} = {\frac{{ik_{3}}} {{\kappa ^{2}}}}u_{x} e^{ik_{3} z},\,\,\,E_{2} = 
{\frac{{ik_{3}}} {{\kappa ^{2}}}}u_{y} e^{ik_{3} z},\,\,\,E_{3} = ue^{ik_{3} 
z},
\end{equation}

\begin{equation}
\label{eq15}
H_{1} = {\frac{{i\omega \varepsilon}} {{\kappa ^{2}}}}u_{y} e^{ik_{3} 
z},\,\,\,H_{2} = {\frac{{i\omega \varepsilon}} {{\kappa ^{2}}}}u_{x} 
e^{ik_{3} z},\,\,\,H_{3} = 0,
\end{equation}

\noindent 
where $u_{x} : = {\frac{{\partial u}}{{\partial x}}}$, $u_{y} 
$ is defined similarly, and $u = u(\hat {x}) = u(x,y)$ solves scalar 
two-dimensional problem (\ref{eq11})-(\ref{eq13}). It is proven in [12] 
that such a problem has a unique solution.

In [19] an asymptotic formula for this solution is derived as $a \to 0$. The 
results consist of the formulas for the solution to the scattering 
problem, of the
equation for the effective field in the new medium obtained by 
embedding many thin perfectly conducting cylinders in the original 
homogeneous medium, characterized by the refraction coefficient 
$n_0^2$,  and of a formula for the 
refraction coefficient in the new medium. This  formula shows 
that by choosing a suitable distribution of thin  perfectly 
conducting cylinders, one can change the refraction 
coefficient, namely, one can make it smaller than $n_0^2$, and
even negative.

The paper is organized as follows.

In Section 2 we derive a linear algebraic system (LAS) for finding some 
numbers that define the solution to problem (\ref{eq11})-(\ref{eq13}) 
with $M > 1$, where $M$ is number of cylinders. 
This is a new feature of our method: instead of looking for some unknown 
boundary functions (currents) we look for just numbers. This method is 
justified only if the cylinders are thin.
We  also derive 
an integral equation for the effective (self-consistent) field in the 
medium with $M=M(a)$ cylinders, $M(a) \to \infty $ as $a 
\to 0$. At the end of 
Section 2 these results are applied to the problem of creating a new 
medium with negative
refraction coefficient  by embedding many thin 
perfectly conducting cylinders into the original (initial) medium.

In Section 3 the numerical results  are
presented. They demonstrate the validity and numerical efficiency 
of the proposed asymptotic method for  solving wave 
scattering problems.
The relative error of the solution to the LAS,
to which the wave scattering problem is reduced, 
is investigated;  the optimal 
parameters $M$, $a$, and $d$
that minimize the error of the 
asymptotic solution of the scattering problem are found. It is 
demonstrated numerically how the 
refraction coefficient of the new medium depends on the 
parameters $M$, $a$, and $d$.

In Section 4 the conclusions are formulated.

\bigskip

%%%%%%%%%%%%%%%%%%%%%%%%%%%%%%%%%%%%%%%%%%%%%%%%%%%
{\bf 2. EM wave scattering by many thin cylinders}
%%%%%%%%%%%%%%%%%%%%%%%%%%%%%%%%%%%%%%%%%%%%%%%%%%%

\bigskip

In this Section, we derive LAS for the numbers 
$u_{e} (\hat {x}_{j} )$. These numbers  
determine the solution of the scattering problem by a rigorous 
asymptotic 
formula. Furthermore, we 
derive an integral equation for the limiting effective field,  
and obtain a simple explicit formula for the refraction coefficient 
$n^{2}$ of 
the new (limiting) medium.

\bigskip

%%%%%%%%%%%%%%%%%%%%%%%%%%%%%%%%%%%%%%%%%%%%%%%%%%%
2.1. Asymptotic formulas for the effective field

\bigskip

Let us assume that the domain $D $ is a union of many small domains 
$D_{m} ,\,\,D = {\bigcup\limits_{m = 1}^{M} {D_{m}}}  $. We assume 
for simplicity that 
$D_{m} $ is a circle of radius $a$ centered at the point $\hat {x}_{m} 
$, and look for the solution to problem (11)-(13) of the form

\begin{equation}
\label{eq16}
u(\hat {x}) = u_{0} (\hat {x}) + {\sum\limits_{m = 1}^{M} 
{{\int\limits_{S_{m}}  {g(\hat {x},t)\sigma _{m} (t)dt}}} },
\end{equation}
where $S_m$ is the boundary of $D_m$, and $dt$ is the element of the 
arclength of  $S_m$.
\noindent

The distribution of the points $\hat {x}_{m} $  in a 
bounded 
domain $\Omega$ on the plane $P = xOy$ is given by formula (\ref{eq2}). 
The incident field is $u_{0} (\hat {x}): = e^{ikx_{2}} $, and 

\begin{equation}
\label{eq17}
g(\hat {x},t): = {\frac{{i}}{{4}}}H_{0}^{(1)} (\kappa \vert \hat {x} - 
t\vert ).
\end{equation}

\noindent
The effective field acting on the $D_{j} $ is defined by the formula

\begin{equation}
\label{eq18}
u_{e} = u_{e}^{(j)} = u(\hat {x}) - {\int\limits_{S_{j}}  {g(\hat 
{x},t)\sigma _{j} (t)dt,\quad \vert \hat {x} - \hat {x}_{j} \vert > a}} ,
\end{equation}

\noindent
or, equivalently,  by the formula

\begin{equation}
\label{eq19}
u_{e} (\hat {x}) = u_{0} (\hat {x}) + {\sum\limits_{m = 1,m \ne j}^{M} 
{{\int\limits_{S_{m}}  {g(\hat {x},t)\sigma _{m} (t)dt}}} } .
\end{equation}

It is  assumed that the distance $d = d(a)$ between neighboring 
cylinders is much greater than $a$:

\begin{equation}
\label{eq20}
d \gg a,\qquad {\mathop {\lim {\frac{{a}}{{d(a)}}}}\limits_{a \to 0}}  = 
0.
\end{equation}

Let us rewrite equation (\ref{eq16}) as

\begin{equation}
\label{eq21}
u = u_{0} + {\sum\limits_{m = 1}^{M} {g(\hat {x},\hat {x}_{m} )Q_{m}}}   + 
{\sum\limits_{m = 1,m \ne j}^{M} {{\int\limits_{S_{m}}  {[g(\hat {x},t) - 
g(\hat {x},\hat {x}_{m} )]\sigma _{m} (t)dt}}} } ,
\end{equation}

\noindent
where

\begin{equation}
\label{eq22}
Q_{m} : = {\int\limits_{S_{m}}  {\sigma _{m} (t)dt}} .
\end{equation}

It was proved in [19] that the second sum in (\ref{eq21}) is negligible 
compared with 
the first one as $a \to 0$. The asymptotic formula for the numbers 
$Q_{m} $ is derived in [19]:

\begin{equation} 
\label{eq23} Q_{m} = {\frac{{ - 2\pi u_{e} (x_{m} 
)}}{{\ln {\frac{{1}}{{a}}}}}}[1 + o(1)],\,\,a \to 0. 
\end{equation}

\noindent
The new idea of our method consists of finding numbers $Q_m$
rather than unknown boundary functions $\sigma_m(t)$. This
leads to a huge gain in the numerical efficiency of our method, and
does not lead to the loss of its accuracy because $a$ is small.
From formulas (21) and (23), one obtains the solution to problem 
(11)-(13) of the form, which is asymptotically, as $a\to 0$,  exact:

\begin{equation}
\label{eq24}
u(\hat {x}) = u_{0} (\hat {x}) - {\frac{{2\pi}} {{\ln 
{\frac{{1}}{{a}}}}}}{\sum\limits_{m = 1}^{M} {g(\hat {x},\hat {x}_{m} )u_{e} 
(\hat {x}_{m} )}}  + o(1).
\end{equation}

\noindent
The numbers $u_{e} (\hat {x}_{m} )$, $1 \le m \le M$, in (24) are not known.
Below we derive  LAS (25) and (31) for finding
 $u_{e} (\hat {x}_{m} )$.  The LAS (31) is of much lower order than 
the LAS (25), and can be interpreted as a collocation method for solving
the integral equation (30) for the self-consistent (limiting) field
in the new medium. The LAS (25), on the other hand,  has a clear 
physical meaning.

Setting $\hat {x} = \hat {x}_{j} $ in  (\ref{eq24}), neglecting $o(1)$ 
term, and 
using the definition (\ref{eq19}) of the effective field, one gets a LAS
for finding the numbers $Q_{m} $:

\begin{equation}
\label{eq25}
u_{e} (\hat {x}_{j} ) = u_{0} (\hat {x}_{j} ) - {\frac{{2\pi}} {{\ln 
{\frac{{1}}{{a}}}}}}{\sum\limits_{m = 1,\,m \ne j}^{M} {g(\hat {x}_{j} ,\hat 
{x}_{m} )u_{e} (\hat {x}_{m} )}} ,\,\,\,1 \le j \le M.
\end{equation}

\noindent
This system can be easily solved numerically if the number $M$ is not 
very large, 
say $M < O(10^{3})$.

\bigskip
%%%%%%%%%%%%%%%%%%%%%%%%%%%%%%%%%%%%%%%%%%%%%%%%%%
2.2. Integral equation for the limiting effective field

\bigskip

If $M$ is very large, $M = M(a) \to \infty ,\,\,a \to 0$, a 
linear integral equation for the limiting effective field in the 
new medium, 
obtained by embedding many thin perfectly conducting  cylinders, is 
derived in 
[19].

Passing to the limit $a \to 0$ in system (25) is done as in [16]. 
Consider a partition of the plane domain $\Omega $, in which the small 
discs $D_m$ are distributed,    into a union of $P$ small 
squares 
$\Delta _{p} $, of size $b = b(a)$, $a\ll b \ll d$. For example, 
one 
may choose $b = O(a^{1 / 4})$, $d = O(a^{1 / 2})$, so that there are 
many discs $D_{m} $ in the square $\Delta _{p} $. We assume that squares 
$\Delta _{p} $ and $\Delta _{q} $ do not have common interior points if 
$p \ne q$. Let $\hat {y}_{p} $ be the center of $\Delta _{p} $. One can 
also choose as $\hat {y}_{p} $ any point $\hat {x}_{m} $ in a domain 
$D_{m} \subset \Delta _{p} $. Since $u_{e} $ is a continuous function, 
one may approximate $u_{e} (\hat {x}_{m} )$ by $u_{e} (\hat {y}_{p} )$, 
provided that $\hat {x}_{m} \subset \Delta _{p} $.

The error of this approximation is $o(1)$ as $a \to 0$. Let us rewrite 
the sum in (25) as follows:

\begin{equation}
\label{eq26}
\frac{2\pi} {\ln (\frac{1}{a})}{\sum_{m = 1,\,m \neq j}^{M} 
{g(\hat {x}_{j} ,\hat {x}_{m} )u_{e} (\hat {x}_{m} )}}  = 2\pi 
{\sum_{p = 1, p\neq q}^{P} {g(\hat {y}_{q} ,\hat {y}_{p} )u_{e} (\hat 
{y}_{p} )} 
}{\frac{{1}}{{\ln {\frac{{1}}{{a}}}}}}\sum_{\hat{x}_{m} \in 
\Delta_{p}}  1,
\end{equation}

\noindent
and use formula (\ref{eq2}) in the form

\begin{equation}
\label{eq27}
{\frac{{1}}{{\ln {\frac{{1}}{{a}}}}}}{\sum\limits_{\hat{x}_{m} \in \Delta 
_{p}}  
{1}}  = N(y_{p} )\vert \Delta _{p} \vert [1 + o(1)].
\end{equation}

\noindent
Here $\vert \Delta _{p} \vert $ is the area of the square $\Delta _{p} $.

From (\ref{eq26}) and (\ref{eq27}) one obtains:

% NEW
\begin{equation}
\label{eq28}
{\frac{{2\pi}} {{\ln {\frac{{1}}{{a}}}}}}{\sum\limits_{\,m \ne j}^{} {g(\hat 
{x}_{j} ,\hat {x}_{m} )u_{e} (\hat {x}_{m} )}}  = 
{\sum_{p = 1, p\neq q}^{P} g(\hat {y}_{q} ,\hat {y}_{p} )N(\hat {y}_{p} 
)u_{e}(\hat {y}_{p} )} \vert \Delta _{p} \vert [1 + o(1)].
\end{equation}

\noindent
The sum in the right-hand side of (\ref{eq28}) is the Riemannian 
sum for the integral

% NEW !!!
\begin{equation}
\label{eq29}
{\mathop {\lim} \limits_{a \to 0}} {\sum\limits_{p = 1}^{P} {g(\hat {y}_{q} 
,\hat {y}_{p} )N(\hat {y}_{p} )u_{e} (\hat {y}_{p} )}} \vert \Delta _{p} 
\vert = {\int\limits_{\Omega}  {g(\hat {x},\hat {y})N(\hat {y})u(\hat 
{y})d\hat {y}}},
\end{equation}

\noindent
where $u(\hat {x}) = {\mathop {\lim} \limits_{a \to 0}} u_{e}(\hat {x})$.
Therefore, system (\ref{eq25}) in the limit $a \to 0$ yields 
the following integral equation for the limiting effective 
(self-consistent) field
$u(\hat {x})$:

\begin{equation}
\label{eq30}
u(\hat {x}) = u_{0} (\hat {x}) - 2\pi {\int\limits_{\Omega}  {g(\hat 
{x},\hat {y})N(\hat {y})u(\hat {y})d\hat {y}}} .
\end{equation}

One obtains a LAS for finding unknown
quantities $u(\hat{y}_q)$, $q=1,2,.....P$,  see 
equation (\ref{eq31}) 
below, if one solves equation (\ref{eq30}) by a 
collocation method with piecewise-constant basis functions. 
Convergence of this method to the unique solution of equation (\ref{eq30}) 
is proved in [13]. Existence and uniqueness of the solution to equation (\ref{eq30}) are 
proved as in [20], where a three-dimensional analog of this equation was 
studied.

The LAS (\ref{eq31}) is used for the numerical calculation of 
the limiting 
effective field and for a comparison of this solution  with the solution 
to LAS (\ref{eq25}), whose order is much larger. 
The LAS is of the form:

\begin{equation}
\label{eq31}
u(\hat {y}_{q} ) = u_{0} (\hat {y}_{q} ) - 2\pi {\sum\limits_{p = 1,\,p 
\ne 
q}^{P} {g(\hat {y}_{q} ,\hat {y}_{p} )N(\hat {y}_{p} )}} u(\hat {y}_{p} 
)|\Delta_p|,\,\,\,1 \le q \le P.
\end{equation}

\noindent 
Comparing the solution to 
(\ref{eq25}) with the solution to LAS (\ref{eq31}) one finds 
the range of applicability of the asymptotic formula (\ref{eq24}) for 
the effective field.

\bigskip
%%%%%%%%%%%%%%%%%%%%%%%%%%%%%%%%%%%%%%%%%%%%%%%%%%
2.3. The refraction coefficient for the new medium

\bigskip

Applying the operator $\Delta _{2} + \kappa ^{2}$ to equation 
(\ref{eq30}) yields the following differential equation for $u(\hat 
{x})$:

\begin{equation}
\label{eq32}
\Delta _{2} u(\hat {x}) + \kappa ^{2}u(\hat {x}) - 2\pi N(\hat {x})u(\hat 
{x}) = 0.
\end{equation}

\noindent
This is a Shr\"{o}dinger-type equation, and $u(\hat {x})$ is the 
scattering 
solution corresponding to the incident wave $u_{0} = e^{i\kappa y}$.

If one assumes that $N(x) = N$ is a constant, then it follows from 
(\ref{eq32}) that the new (limiting) medium, obtained by embedding many 
perfectly conducting circular cylinders, has new parameter  
$\kappa _{N}^{2} : = \kappa ^{2} - 2\pi N$. 
This means that $k^{2} = \kappa ^{2} - k_{3}^{2} $ is replaced by 
$\bar {k}^{2} = k^{2} - 2\pi N$. The 
quantity $k_{3}^{2} $ is not changed. One has $\bar {k}^{2} = \omega 
^{2}n^{2},\,\,k^{2} = \omega ^{2}n_{0}^{2} $. Consequently, $n^{2} / 
n_{0}^{2} = (k^{2} - 2\pi N) / k^{2}$. Therefore, the new refraction 
coefficient $n^{2}$ is

\begin{equation}
\label{eq33}
n^{2} = n_{0}^{2} (1 - 2\pi Nk^{ - 2}).
\end{equation}

\noindent
Since the number $N > 0$ is at our disposal, equation (\ref{eq33}) shows that 
choosing suitable $N$ one can create a medium with a smaller than $n_{0}^{2}$, 
refraction coefficient, even with negative refraction coefficient $n^2$.

In practice one does not go to the limit $a \to 0$, but chooses a 
sufficiently small $a$. As a result, one obtains a medium with a 
refraction 
coefficient $n_{a}^{2} $, which differs from (\ref{eq33}), but the error 
tends to zero as $a\to 0$, and one has 
${\mathop {\lim 
}\limits_{a \to 0}} n_{a}^{2} = n^{2}$.

\bigskip
%%%%%%%%%%%%%%%%%%%%%%%%%%%%%%%%%%%%%%%%%%%%%
{\bf 3. Numerical results}
%%%%%%%%%%%%%%%%%%%%%%%%%%%%%%%%%%%%%%%%%%%%%

\bigskip

Some  algorithms for computational modeling of the wave scattering by 
many  small 
particles were developed in [1] for the acoustic wave scattering and 
generalized in [2], [3] for electromagnetic (EM) wave scattering. It was 
proved in these papers that the asymptotic solution of the many-body 
wave scattering problem, proposed in [10]-[16],[20], is computationally 
efficient and yields accurate numerical results. On the basis of
this theory a recipe for creating a material with a desired refraction 
coefficient was formulated. This recipe was verified numerically in [2]. 

In this Section, numerical results are presented. These results  
demonstrate
the efficiency of the asymptotical approach for solving the EM wave 
scattering 
problem in the media with many embedded perfectly conducting 
cylinders of small radius $a$, and the possibility 
to create the medium with a negative refraction coefficient. 

The first portion of the numerical results demonstrates the 
approximation errors of 
the numerical solutions to the LAS (\ref{eq25}) compared to 
 the numerical solution of LAS 
(\ref{eq31}), corresponding to the collocation method for solving of 
the integral 
equation (\ref{eq30}) for the limiting field. The rest of the 
numerical results demonstrate the 
possibility to 
create the media with  negative refraction coefficient, and 
yields  optimal parameters  $M$, $a$, and $d$, for creating
such a coefficient.

The following numerical problems are important from the practical point 
of view:

- to determine the values of the parameters  
$M$, $a$, and $d$, that provide the solution to LAS (\ref{eq25}) 
with the desired 
accuracy, for example, with relative error of the order $10^{ - 3} - 
10^{ - 4})$;

- to investigate the convergence of the LAS (\ref{eq31}) and to 
determine 
the optimal values of the parameters  $M$, $a$, and $d$,  which provide 
such convergence;

- to compare the solution to LAS (\ref{eq25}) and LAS (\ref{eq31}) and 
to find the range of  $M$, $a$, and $d$ that provide accurate 
solutions;

- to determine the values of  $M$, $a$, and $d$, which yield  
negative refraction coefficient.

\bigskip

%%%%%%%%%%%%%%%%%%%%%%%%%%%%%%%%%%%%%%%%%%%%%%%%%%%%%%%%5
3.1.  The accuracy of the solution to LAS (\ref{eq25})

\bigskip

The numerical procedure for checking the accuracy to LAS (\ref{eq25}) 
consists in calculations with various values of parameters $M$, $a$, and 
$d$ at $k=1.41$. First, we study the convergence of the solution depending on 
the number $M$ of embedded into $\Omega $ cylinders. The radius $a$ of 
the cylinders is changed, the distance $d$  between 
neighboring cylinders is kept in the range $d\ge 10a$. 
In Fig. 1, the relative errors of the solutions to (\ref{eq25}) 
%($\vert u\vert$ is taken into account for consideration here and throughout 
%this section) 
are shown for the case when number $M$ of cylinders grows from 
25 to 3200. The values of $\sqrt {M} $ are depicted along the $x$ axis. 
Because the exact solution to (\ref{eq25}) is not known, the relative 
error was calculated by formula

\begin{equation}
\label{eq34}
{\rm R}{\rm E} = {\frac{{(\vert u_{2M} \vert - \vert u_{M} \vert )}}{{\vert 
u_{2M} \vert}} }
\end{equation}

\noindent
instead of generally used

\begin{equation}
\label{eq35}
{\rm R}{\rm E} = {\frac{{(\vert u_{\ast}  \vert - \vert u_{M} \vert 
)}}{{\vert u_{\ast}  \vert}} },
\end{equation}

\noindent 
where $u_{2M} $ and $u_{M} $ are the solution to (\ref{eq25}) 
with $2M$ and $M$ cylinders respectively, $u_{\ast} $ is exact solution.

The maximal error is observed for $\sqrt {M} = 5$ and it is equal to $11.2\%$. 
The error diminishes when $M$ grows and its value is $0.2\% $ at $M =1600$. 
The error practically does not depend on the radius $a$ of cylinder 
for big $M$. The curves presented in Fig. 2 show that absolute error has 
similar character and its minimal value is $0.002$ for considered $a$.

The above numerical results are obtained for  
$d\ge 10a$. Computationally, the values of the error depend on this 
ratio. 
This can be used for minimization of the error by choosing different $a$ and 
$d$. The numerical calculations show that there is an optimal value of 
the distance $d$ between cylinders, which provide the minimal error if  $a$ is fixed. 
This optimal distance $d$ depends on the number $M$ of cylinders in 
$D$ and varies in the range $14.5a - 40a$.

\begin{figure}[tbh]
\includegraphics[width=0.5\textwidth]{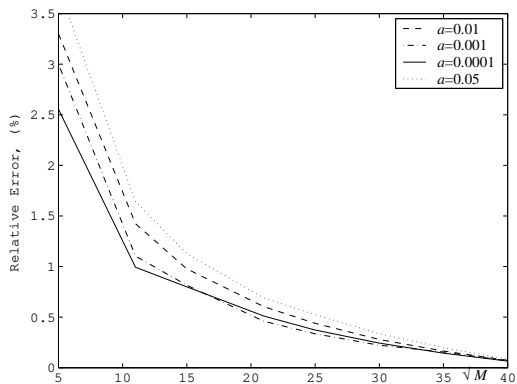}
\caption{Relative error of solution to (25) versus $M$, $k=1.41$}
\label{fig1}
\end{figure}

\begin{figure}[tbh]
\includegraphics[width=0.5\textwidth]{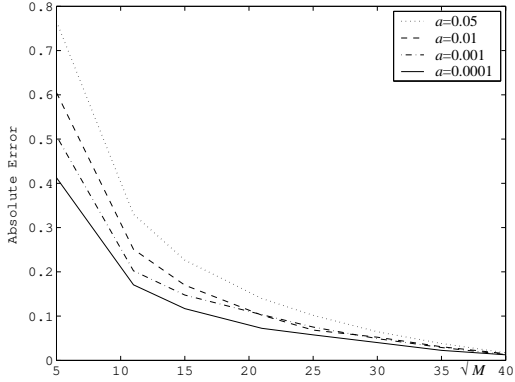}
\caption{Absolute error of solution to (25) versus $M$, $k=1.41$}
\label{fig2}
\end{figure}

Note that the errors of $E$ components depend  on 
the ratio of $k$ and $k_{3} $. The numerical results show that values of 
$\kappa $ ($\kappa ^{2} = k^{2} - k_{3}^{2} )$ in a neighborhood of 1 
yield the minimal errors. This implies $k_{3}=1.0$ on account of $k^{2}=(1.41)^{2}=2$.

\bigskip
%%%%%%%%%%%%%%%%%%%%%%%%%%%%%%%%%%%%%%%%%%%%%%%%%%%%%%%
3.2. The relative error for the solution to LAS (\ref{eq31})

\bigskip

The collocation method [13] for solving LAS (\ref{eq31}), 
corresponding to the limiting equation (\ref{eq30}), is applied to check 
the accuracy of the numerical solution of LAS (\ref{eq31}). The relative error is 
defined in the previous subsection. In Fig. 3 the dependence of the relative 
error on the number $P$ of the collocation points is shown. When $P$ 
is small, for example, $P = 25$, this error is large: it is equal to 
$30.2\% $, $27.4\% $, 
and $10.4\% $ for $a = 0.05$, $a = 0.01$, and $a = 0.001$, respectively; 
the error is equal to $2\% $ for $a = 0.0001$. In the considered range 
of $P$, the error depends on $a$. 
%%% - ??What are $k$ and $ka$ here??
The value of $ka$ does not exceed $0.0705$ here.
The smallest value of $a = 0.0001$ provides low 
error for all $P$. The values of absolute error are shown in Fig. 4. The 
maximal value of this error at $\sqrt{\it M}=35$ does not exceed $0.005$ 
and  is achieved at $a=0.05$.

\begin{figure}[tbh]
\includegraphics[width=0.5\textwidth]{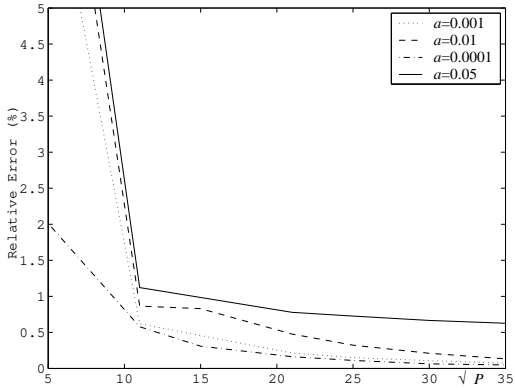}
\caption{Relative error of solution to (\ref{eq31}) versus $P$}
\label{fig3}
\end{figure}

\begin{figure}[tbh]
\includegraphics[width=0.5\textwidth]{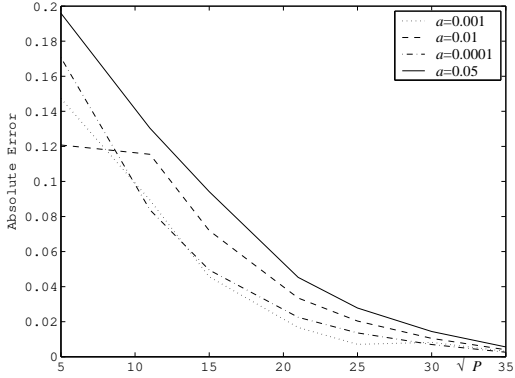}
\caption{Absolute error of solution to (\ref{eq31}) versus $P$}
\label{fig4}
\end{figure}

The values of the errors for the field $u$ and its $E$ components for 
the large 
$P$ at $a = 0.05$ are shown in Table 1. The values of $N(x)=N$, with 
various constant values of $N$, 
are calculated by formula (\ref{eq2})
% assuming the uniform 
%distribution of cylinders in the domain $\Omega$, are 
and are shown in the last column of Table 1.

\begin{center} Table 1. \end{center} 
\hrule 
\begin{center} 
\begin{tabular}{lllllll} $\sqrt {P} $ & Rel.Err. & Abs.Err. & Abs.Err. & 
Abs.Err. & Abs.Err.  &$ N$ \\ $\quad$ & of $\vert u\vert $ & of 
$\vert u\vert $ & of $\vert E_{1} \vert $ & of $\vert E_{2} \vert $ & of 
$\vert E_{3} \vert $ & $\quad$ \\ 
% \hrule 
40 & 0.76\% & 0.0041 & 0.0185& 0.0205 & 0.0163 & 61.53 \\
 45 & 0.54\% & 0.0034 & 0.0166 &0.0183 & 0.0147 & 77.89 \\ 
50 & 0.36\% & 0.0028 & 0.0142 & 0.0165 & 0.0134 & 96.16 \\ 
55 & 0.22\% & 0.0024 & 0.0126 & 0.0150 & 0.0123 & 116.36 \\
 60 & 0.10\% & 0.0020 & 0.0113 & 0.0139 & 0.0114 & 138.47 \end{tabular} 
\end{center} 
\hrule
\medskip

The above calculations were carried out at the values of $N$ that 
were determined by formula (\ref{eq2}) for the prescribed $M$ and $d$. 
This implies the following formula

\begin{equation}
\label{eq36}
 N = {\frac{{{\mathcal N}(\Omega)}}{{\ln (1 / a)\vert \Delta _{\Omega} 
\vert}} },
\end{equation}

\noindent
which agrees with formula (43) in [19] when $N(x)=N={\rm const}$, the 
distribution of particles is assumed in the whole $\Omega$, and $\Omega$ 
is the union of the non-intersecting domains $\Delta _{p} $.

In formula (\ref{eq36}) the quantity  ${\mathcal N}(\Omega)$ is the 
total number of the embedded 
cylinders in 
$\Omega$, $\vert \Delta _{\Omega} \vert $ is the area of $\Omega$, $k=1.41$. The 
calculations show 
that the  value of $N(x)$, calculated by formula  (\ref{eq36}),
can be varied so that it will provide the minimal error for 
the solution to LAS (\ref{eq31}). In Fig. 5, the relative and 
absolute 
errors for the solution to LAS (\ref{eq31}) are shown in a neighborhood 
of various values of
$N$, calculated by formula (\ref{eq36}). The first vertical line at the 
$x$ 
axis corresponds to $N = 6.49$, calculated for $a = 0.0001$, and the 
second one corresponds to $N = 7.86$ for $a = 0.0005$. The minimal 
values of the  errors are to the  left of the values of $N$, calculated by 
(\ref{eq36}).
% and are less several times. 
For the considered parameters, 
the relative error decays from $0.72\%$ to $0.31\%$ for $a = 
0.0005$, and it decays from $0.64\%$ to $0.15\%$ for $a = 0.0001$. 
The absolute error decays from $0.05$ to $0.009$ and from $0.025$ to 
$0.005$, respectively.

\begin{figure}[tbh]
\includegraphics[width=0.5\textwidth]{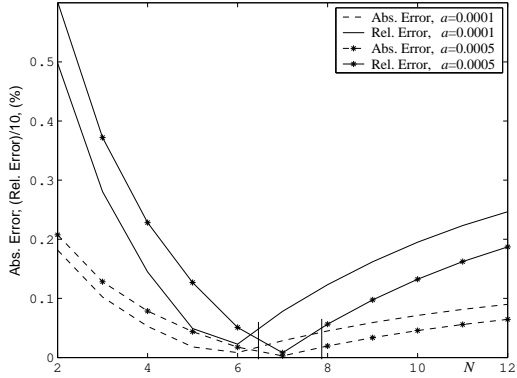}
\caption{The absolute and relative errors of the field versus $N$}
\label{fig5}
\end{figure}

\begin{figure}[tbh]
\includegraphics[width=0.5\textwidth]{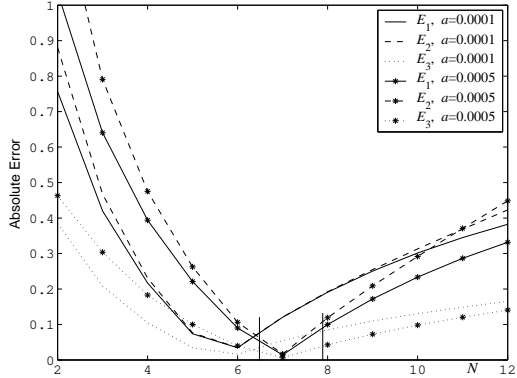}
\caption{The absolute errors of $E$ components versus $N$}
\label{fig6}
\end{figure}

The absolute error for the $E_j$, $j=1,2,3,$ components of the field is 
presented in Fig. 6. The 
errors for $E_1$ and $E_2$ are higher than that for $E_3$ 
because the components $E_{1} $ and $E_{2} $ contain the derivative of 
function $H_{0}^{(1)} (kr)$ at the small values of $kr$, while 
$E_{3}$ does not contain the derivative.

The value of $N$ plays a role of an additional parameter varying which 
one can decrease the error. The value of $N$ can be changed 
by changing the distance $d$ 
between neighboring cylinders while keeping fixed
their number in the area where $N$ is being changed.

\bigskip
%%%%%%%%%%%%%%%%%%%%%%%%%%%%%%%%%%%%%%%%%%%%%%%%%%%%%
3.3. Comparison of solutions to LAS (\ref{eq25}) and LAS (\ref{eq31})

\bigskip

The accuracy of the asymptotic formula (24) was investigated by 
comparing the solutions to LAS (\ref{eq25}) and to LAS (\ref{eq31}). The 
solution to (\ref{eq31}) with $P = 4900$ collocation points is 
considered the {\it benchmark solution}, $k=1.41$. The relative error of this 
solution does not exceed $1\% $ at the considered values of $a$. This 
error is maximal at $a = 0.05$ and it decays if $a$ decreases.

In Fig. 7 the relative and absolute errors of the solution to LAS (25) 
are shown at various $a$. The maximal value of the relative error is 
observed at 
$M = 25$ and it is equal to $32.7\% $, $27.3\% $, and $16.6\% $ at $a = 
0.01$, $a = 0.001$, and $a = 0.0001$ respectively. This error for $\sqrt 
{M} = 35$ is equal to $2.2\% $, $1.9\% $, and $2.4\% $.

The absolute error for $E_1$ and $E_2$ components is shown in Fig. 8. As 
in the preceding subsection (see Fig. 6) this error is  higher than 
the error for $E_3$. The largest error for $E_{2} $ is equal to $1.2$ for 
$a = 0.01$ when $M = 25$; the minimal value of the error for this $M$ is 
obtained for $E_{3} $ at $a = 0.0001$, and is equal to $0.13$. 
The minimal value of the error for $E_{3}$ component
 is obtained  when $\sqrt {M} = 35$ and 
this error is  $0.02$.

\begin{figure}[tbh]
\includegraphics[width=0.5\textwidth]{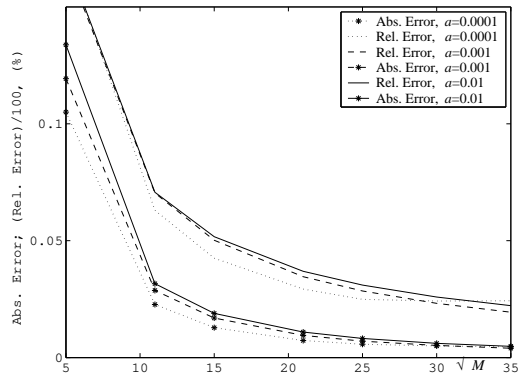}
\caption{Absolute and relative errors of $u$ versus number $M$ of cylinders, $k=1.41$}
\label{fig7}
\end{figure}

\begin{figure}[tbh]
\includegraphics[width=0.5\textwidth]{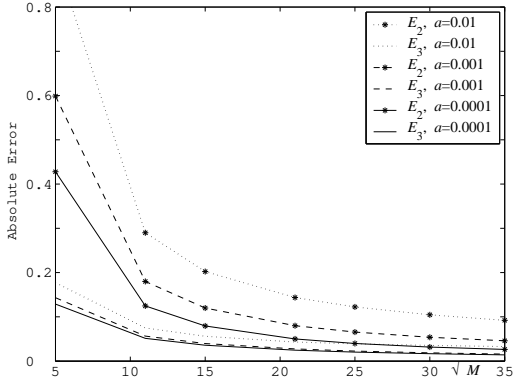}
\caption{Absolute error of $E$ components versus number $M$ of cylinders, $k=1.41$}
\label{fig8}
\end{figure}

The above results are obtained in the case when the distance $d$ 
between the cylinders is fixed. It turns out that the distance parameter 
$d$ influences also the error of the solution to equation (24) when the number 
of the cylinders is fixed. The errors of 
the solution to equation  (25) when $M = 900$ for various values of  $d$
are shown in Fig. 9 and Fig. 10. 
The benchmark 
solution to LAS (\ref{eq30}) is the same as in the preceding example.
There is an optimal value of $d$, which provides the minimal 
value of the error. The values of $na$, $n=1,2,3,....$, 
are shown
along the $x$-axis. The minimal error equals to $0.57\% $ 
and is obtained when 
$a = 0.01$ and $na = 12.6$; it is equal to $0.29\% $ when $a = 0.001$ and 
$na = 16.5$, and it is equal to $0.23\% $ when $a = 0.0001$ and $na =21.6$. 
The minimal values of absolute error when $a = 0.001$ and $a = 0.0001$ are 
shifted to the left in comparison to the minimal relative error.

\bigskip
\begin{figure}[tbh]
\includegraphics[width=0.5\textwidth]{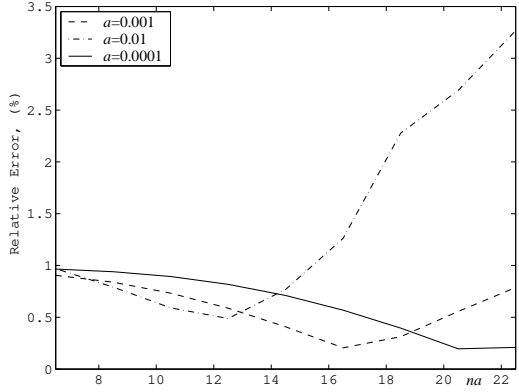}
\caption{Relative error of $u$ versus distance $d$ between cylinders, $k=1.41$}
\label{fig9}
\end{figure}

\begin{figure}[tbh]
\includegraphics[width=0.5\textwidth]{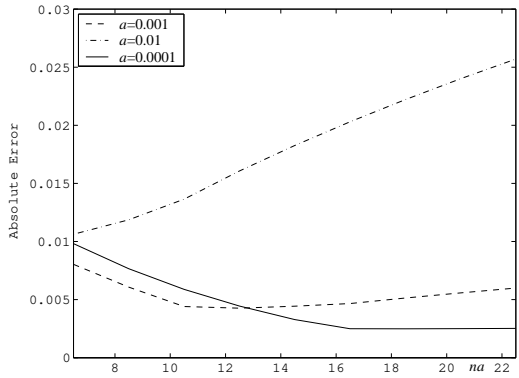}
\caption{Absolute error of $u$ versus distance $d$ between cylinders, $k=1.41$}
\label{fig10}
\end{figure}

\bigskip

%%%%%%%%%%%%%%%%%%%%%%%%%%%%%%%%%%%%%%%%%%%%%%%%%%%%%%%%%%%%%%
3.4. The refraction coefficient of the new medium

\bigskip

One can conclude from formula (\ref{eq33}) that the value of the 
refraction 
coefficient $n^{2}$ depends on the wave number $k$, 
on the parameter $N$, on  $a$,  
on $M$, and on $d$.
In Fig. 11 and Fig. 12  the dependence of $n^{2}$ on 
%the wave number 
$k$ is shown for two various values of $d$. The number $M$ of 
the cylinders is 
equal to 225. The cylinders are placed equidistantly in a square 15 
cylinders on the side of the square.
%%% - ?? What is the size of this square? 
The lengths $l$ of square are equal to $0.1386m$  and  $0.2857m$
in the Fig. 11 and Fig. 12 respectively.
%% - What is the value of $d$?  ??
Consequently, the values of $d$ are equal to $0.0099m$ and 
$0.0204m$.

The value of $k$ has dimension $L^{-1}$, where $L$ is length, the  
$a$ and $d$ have dimension $L$, and the values $n^{2}$ of the refraction 
coefficient are normalized to the value $11.1254 \cdot 10^{ - 
18}{\frac{{{\rm s}{\rm e}{\rm k}^{2}}}{{{\rm m}^{2}}}}$.
This value is obtained by multiplying $\varepsilon_{0}=8.85 \times 
10^{-12}  F/m$
and $\mu_{0}=4\pi \times 10^{-7} H/m$, taking into account the formula 
$n_{0}^{2}=\varepsilon_{0}\mu_{0}$, where $F$ stands for farad, and $H$
stands for henry, $[H]=\frac {T^2}{F}$, $[\cdot]$ stands for the 
dimension of a physical quantity, and $T$ stands for time. 

At the smaller $d =0.0099 m$ (see Fig. 11) 
the values of $n^{2}$ differ considerably from the refraction coefficient 
$n_{0}^{2} = 1$ 
of initial media, because more cylinders are embedded per unit area. 
It is seen from Fig. 12 that the refraction coefficient $n^{2}$ 
when $d = 0.0204m$ is close to $n_{0}^{2} $. An increase of $k$ 
forces $n^{2}$ to get closer to initial refraction coefficient  $n_{0}^{2} $. 
This is observed for all considered values of $a$.
The considered values of the $M$, $a$, and $d$  yield the ratio 
$\frac {a} {d}$ less than 0.05, so  condition (\ref{eq20}) is 
satisfied.

\begin{figure}[tbh]
 \includegraphics[width=0.5\textwidth]{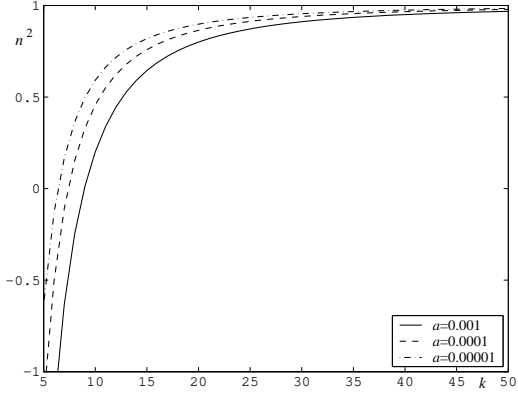} 
\caption{Values of $n^{2}$ versus wave number $k$, $M=225$, 
$d = 0.0099m$} \label{fig11} \end{figure}

\begin{figure}[tbh]
\includegraphics[width=0.5\textwidth]{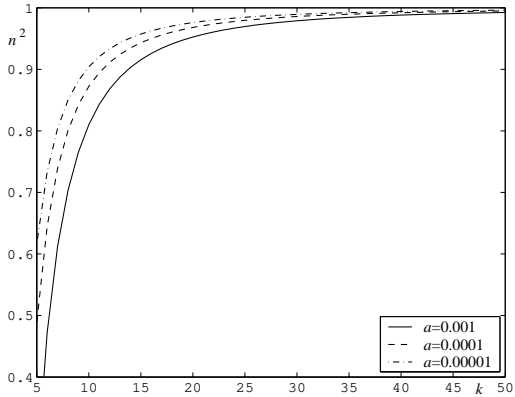}
\caption{Values of $n^{2}$ versus wave number $k$, $M=225$, $d = 0.0204m$}
\label{fig12}
\end{figure}

%\bigskip

The numerical results presented in Fig. 13 demonstrate a possibility to 
create 
the medium with various refraction coefficients $n^{2}$ depending on the 
distance $d$ between the
cylinders, when $a$ and $k$ are fixed. The results are shown for $a = 
0.0001$ 
and $k = 20.0$. At the small values of $M$ the values of $n^{2}$ are 
changed considerably, and 
when $M$ increases $n^{2}$ tends to the following values: $n^{2} = - 
0.45$, $n^{2} = 0.06$, $n^{2} 
= 0.35$, 
and $n^{2} = 0.64$ when $d = 20a$, $d = 25a$, $d = 30a$, and $d = 
40a$, 
respectively. Note that at the considered values of the parameters 
the relative error of the solution to LAS (25) does not exceed 
$2.34\% $, 
$1.69\% $, $1.18\% $, and $0.93\% $ for $d = 20a$, $d = 25a$, $d 
= 30a$, and 
$d = 40a$ at $\sqrt {M} = 30$, and the relative error decays when 
$M$ grows.
%% - ?? What are the values of $M$ here and in Fig 11,12?  ?? - inserted
%%% - M is variable here
%% - ?? In Fig 9,10  what is $k$?  - inserted
%% - It is, probably,better to use n ka
%% - ?? rather than na ?? 
%%% since k \neq 1, na is better
%% - ?? In Fig 13 $ n^2=0$ when$ M=0$. This is wrong ?? - M starts from 25 !!!

\begin{figure}[tbh]
\includegraphics[width=0.5\textwidth]{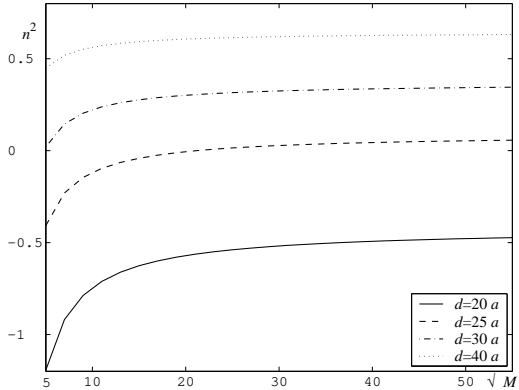}
\caption{Values of $n^{2}$ versus the number $M$ of cylinders}
\label{fig13}
\end{figure}
%\bigskip

Consequently, one can change the refraction coefficient $n^{2}$ 
by changing $k$, $a$, $M$, and $d$. 

\bigskip

{\bf 4. Conclusions}

\bigskip

Asymptotic solution is given for the problem of EM wave 
scattering by many perfectly conducting parallel cylinders of 
small radii $a$, $ka<<1$. 
An equation for the effective (self-consistent) field in the limiting 
medium is obtained 
when $a \to 0$ and the distribution of the embedded cylinders is given 
by formula (\ref{eq2}). The theory yields formula (\ref{eq33}) 
for the refraction coefficient of the new (limiting) medium
obtained by embedding of these cylinders into the initial homogeneous 
medium. This formula 
shows how the distribution of the cylinders influences the refraction 
coefficient.

The numerical results confirm the validity and efficiency of the 
asymptotic method for solving the above scattering problem.
The optimal values of the parameters $k, a, d, M, N$, that
minimize the error of the solution to the scattering problem,  
are found numerically. 
It is shown both theoretically and  numerically that one can create
negative refraction coefficients in the new medium. 

%% can be 

\bigskip

%%%%%%%%%%%%%%%%%%%%%%%%%%%%%%%%%%%%%%%%
{\bf References}

%\bigskip

[1] M. I. Andriychuk  and A. G. Ramm.  Scattering by many small particles and 
creating materials with a desired refraction coefficient. 
Intern. Journ. of Comp. Sci. and Math., Vol. 3, Nos.1/2, (2010),
p. 102-121.

[2] M. I. Andriychuk, A. G. Ramm. Numerical Solution of Many-Body Wave 
Scattering Problem for Small Particles and Creating Materials with Desired 
Refraction Coefficient. In Book: Numerical Simulations of Physical and 
Engineering Processes. Ed. By Jan Awrejcewich, InTech, Rieka, 2011, p. 3-28.

[3] M. I. Andriychuk,  S. W. Indratno, A. G. Ramm. Electromagnetic wave
scattering by a small impedance particle: theory and modeling. Optics
Communications, 285, (2012), p. 1684-1691.

[4] D. R. Denison, R. W. Scharstein. Decomposition of the scattering by a 
finite linear array into periodic and edge components. Microwave and Optical 
Technology Letters, Vol. 9, Issue 6, (1995), p. 338-343.

[5] S. Dubois, A. Michel, J. P. Eymery, J. L. Duvail, and L. Piraux. 
Fabrication and properties of arrays of superconducting nanowires. Journ. of 
Materials Research, 14, (1999), p. 665-671.

[6] L. Landau, E. Lifshitz. Electrodynamics of continuous media. Pergamon 
Press, London, 1984.

[7] P. Martin. Multiple Scattering. Cambridge Univ. Press, Cambridge, 2006.

[8] C. Mei, B. Vernescu. Homogenization methods for multiscale mechanics. 
Word Sci., New Jersey, 2010.

[9] S.-M. Park, G. S. W. Craig, Y.-H. La, H. H. Solak, and P. F. Nealey. 
Square Arrays of Vertical Cylinders of PS-$b$-PMMA on Chemically Nanopatterned 
Surfaces. Macromolecules, 40, (14), (2007), p. 5084-5094.

[10] A. G. Ramm. Distribution of particles which produces a "smart" 
material. Journ. Stat. Phys., 127, No 5, (2007), p. 915-934.

[11] A. G. Ramm. Electromagnetic wave scattering by small bodies. Phys. 
Lett. A, 372/23, (2008), p. 4298-4306.

[12] A. G. Ramm. Wave scattering by many small particles embedded in a 
medium. Phys. Lett. A, 372/17, (2008), p. 3064-3070.

[13] A. G. Ramm. A collocation method for solving integral equations. Int. 
Journ. Comp. Sci. and Math., 3, No 2, (2009), p. 222-228.

[14] A. G. Ramm. A method for creating materials with a desired refraction 
coefficient. Intern. Journ. Mod. Phys. B, 24, (2010), p. 5261-5268.

[15] A. G. Ramm. Materials with desired refraction coefficient can be 
creating by embedding small particles into the given material. Intern. 
Journ. on Struct. Changes in Solids, 2, No 2, (2010), p. 17-23.

[16] A. G. Ramm. Wave scattering by many small bodies and creating materials 
with a desired refraction coefficient. Africa Matematika, 22, No 1 (2011), 
p. 33-55.

[17] A. G. Ramm. Electromagnetic wave scattering by a  small
impedance particle of arbitrary shape. Optics
Communications, 284, (2011), p. 3872-3877.

[18] A. G. Ramm. Scattering of scalar waves by many
small particles. AIP Advances, 1, (2011), p. 022135.

[19] A. G. Ramm. Scattering of electromagnetic waves by many thin cylinders. 
Results in Physics, 1, No 1, (2011), p. 13-16.

[20] A. G. Ramm.  Many body wave scattering by small bodies and 
applications. Journ. Math. Phys., Vol. 48, No. 10, (2007), p. 103511.

[21] A. G. Ramm. Electromagnetic wave scattering by many small
perfectly conducting particles of an arbitrary shape. Optics
Communications, 285, (2012); http://dx.doi.org/10.1016/j.optcom.2012.05.010 

[22] A. F. J. Smith and A. A. Wragg. An electrochemical study of mass 
transfer in free convection at vertical arrays of horizontal cylinders. 
Journ. of Appl. Electrochem., 4, No 3, (2009), p. 219-228.

[23] Q. Zhou and R. W. Knighton. Light scattering and form birefringence of 
parallel cylindrical arrays that represent cellular organelles of the 
retinal nerve fiber layer. Applied Optics, Vol. 36, Issue 10, (1997), p. 
2273-2285.

\end{document}